\documentclass{article}
\usepackage{spconf,amsmath}
\usepackage{graphicx}

\usepackage{amsfonts,color}

\DeclareMathOperator*{\argmin}{arg\,min}

\newcommand*{\fHatTilde}{\hat{\skew{1}\tilde{\mathbf{f}}}}

\title{Graph Spectral Characterization of white matter fMRI data on voxel resolution graphs derived from Diffusion MRI}
\title{Graph Spectral Characterization of fMRI data on voxel resolution gray matter graphs}
\title{Spectral Analysis of fMRI data on voxel-wise gray matter graphs}
\title{Spectral Characterization of functional MRI data on \\ voxel-resolution cortical graphs}
\title{Graph Spectral Characterization of functional MRI data on \\ voxel-resolution cortical graphs}
\title{Spectral Quantification of functional MRI data on \\ voxel-resolution cortical graphs}
\title{Spectral characterization of functional MRI data on \\ voxel-resolution cortical graphs}

\name{Hamid~Behjat$^{1}$, Martin~Larsson$^{2}$
\thanks{*This work was supported by the Swedish Research Council under Grant~2018-06689, and in part by the Royal Physiographic Society of Lund. Correspondence should be addressed to H. Behjat, e-mail: hamid.behjat@bme.lth.se. 
}
} 

\address{
\small $^{1}$ Department of Biomedical Engineering, Lund University, Lund, Sweden
\\ 
\small $^{2}$ Centre for Mathematical Sciences, Lund University, Lund, Sweden}

\begin{document}
%
\maketitle
\begin{abstract}
The human cortical layer exhibits a convoluted morphology that is unique to each individual. Conventional volumetric fMRI processing schemes take for granted the rich information provided by the underlying anatomy. 
We present a method to study fMRI data on subject-specific cerebral hemisphere cortex (CHC) graphs, which encode the cortical morphology at the resolution of voxels in 3-D. Using graph signal processing principles, we study spectral energy metrics associated to fMRI data, on 100 subjects from the Human Connectome Project database, across seven tasks. Experimental results signify the strength of CHC graphs' Laplacian eigenvector bases in capturing subtle spatial patterns specific to different functional loads as well as to sets of experimental conditions within each task. 

\end{abstract}
\section{Introduction}
\label{sec:intro}
Functional magnetic resonance imaging (fMRI) is a key modality in the study of human brain activity based on the blood-oxygen-level-dependent (BOLD) signal. 
Despite the apparent confinement of the BOLD signal to underlying anatomy, the use of anatomically-informed methods for analysing fMRI data is generally sparse. 
By anatomically-informed methods we refer to schemes that aim to enhance the processing of fMRI data by exploiting in one way or another knowledge about the underlying anatomy of the data. 
In this line, a range of schemes have been proposed for various fMRI analyses procedures on the cortical surface, namely, interpolation \cite{Grova2006}, smoothing \cite{Chung2004} and decompositions using cortical encoding spatial bases \cite{Gabay2017}. 
A number of anatomically-informed volumetric schemes have also been proposed, within applications such as bilateral spatial filtering \cite{Rydell2008}, Markov random field regularization \cite{Ou2010} and spatiotemporal fMRI deconvolution \cite{Farouj2017}.
The most relevant research related to the present work are \cite{Behjat2015, Behjat2013}, where voxel-resolution gray matter graphs were designed and exploited for enhanced activation mapping in gray matter, within group level \cite{Behjat2015} and subject level \cite{Behjat2013} analyses. 
Here, we leverage CHC graphs which have superior morphology encoding properties compared to gray matter graph designs in \cite{Behjat2015, Behjat2013}.
Their superiority is due to that they, firstly, use extensively preprocessed FreeSurfer \cite{Fischl2012} extracted ribbon files as their basis, and secondly, leverage a pruning process that remove anatomically unjustifiable graph edges. 
The Laplacian spectra of CHC were recently shown to manifest morphological variations associated to gender as well as hemispheric asymmetry \cite{Maghsadhagh2019}. 
Moreover, the main focus in \cite{Behjat2015, Behjat2013} is to study the performance of the method based on the resulting activation maps, rather than exploring graph spectral properties of fMRI data. 
A thorough understanding of spectral properties of fMRI data on cortical graphs is key for future work to enhance fMRI activation mapping using CHC graphs. 
This paper takes a step in this direction.  
   



\section{Materials and Methods}

\subsection{Graph signal processing fundamentals}
An undirected, non-weighted graph, denoted $\mathcal{G}=(\mathcal{V},\mathcal{E},\mathbf{A})$, consists of a set $\mathcal{V} = \{1,2,\ldots,N\}$ of $N$ vertices and a set $\mathcal{E}$ of edges (i.e., pairs ($i,j$) where $i,j \in \mathcal{V}$), and $\mathbf{A}$ denotes the graph's adjacency matrix. The graph's normalized Laplacian matrix $\mathcal{L}$ is given as $\mathbf{\mathcal{L}} = \mathbf{I}-\mathbf{D}^{-\frac{1}{2}}\mathbf{A}\mathbf{D}^{-\frac{1}{2}}$, where $\mathbf{D}$ is the diagonal matrix of vertex degrees and $\mathbf{I}$ denotes the identity matrix. The Laplacian matrix can be diagonalized as 
\begin{equation}
\mathbf{\mathcal{L}} = \mathbf{U}\mathbf{\Lambda}\mathbf{U}^{*},
\label{eq:L}
\end{equation}
where $\mathbf{\Lambda}$ is the diagonal matrix of eigenvalues $\lambda_{1}, \ldots, \lambda_{N} := \lambda_{\text{max}}$ of $\mathbf{\mathcal{L}}$ and the columns $\mathbf{u}_1, \cdots, \mathbf{u}_{N}$ of $\mathbf{U}$ are the associated eigenvectors of $\mathbf{\mathcal{L}}$; hereon, we refer to eigenvectors as eigenmodes, in line with the convention in the neuroimaging community. The eigenvalues define the Laplacian spectrum of $\mathcal{G}$ \cite{Chung1997}, a space equivalent to the Euclidean Fourier domain. 

Let $\mathbf{f}\in \mathbb{R}^{N}$ denote a graph signal residing on the vertices of $\mathcal{G}$. The \textit{graph Fourier transform} (GFT) of $\mathbf{f}$, denoted $\hat{\mathbf{f}} \in \mathbb{R}^{N}$, is obtained as $\hat{\mathbf{f}} = \mathbf{U}^{*}\mathbf{f}$, which satisfies the Parseval energy conservation relation \cite{Shuman2015acha}: $||\mathbf{f}||_2^{2} = ||\hat{\mathbf{f}}||_2^{2}$. 

A graph filter can be conveniently defined in the graph spectral domain. Given the spectral profile of a filter $h: [0,\lambda_{\text{max}}] \rightarrow \mathbb{R}$, $\mathbf{f}$ can be filtered with $h(\cdot)$ as 
\begin{align}
h(\mathbf{\mathcal{L}}) \mathbf{f} 
= \mathbf{U} h(\mathbf{\Lambda}) \mathbf{U}^{*}\mathbf{f}
= \mathbf{U} h(\mathbf{\Lambda}) \hat{\mathbf{f}}.
\label{eq:filtering-a}
\end{align} 
A shortcoming with this approach to filtering is that it requires $\mathbf{\Lambda}$ and $\mathbf{U}$, i.e., diagonalization of $\mathcal{L}$. This is computationally cumbersome for large graphs, and in particular, for the graphs proposed in this work that are typically of size $~120$ K vertices it is highly impractical. An alternative approach is to exploit a polynomial approximation of the spectral kernel $h(\cdot)$, denoted $\mathcal{P}(h):[0,\lambda_{\text{max}}] \rightarrow \mathbb{R}$, and implement filtering as \cite{Hammond2011} 
\begin{align}
h(\mathbf{\mathcal{L}}) \mathbf{f} 
\stackrel{(\ref{eq:filtering-a})}{=} \mathbf{U} h(\mathbf{\Lambda}) \mathbf{U}^{*}\mathbf{f}
\approx 
\mathbf{U}  \mathcal{P}(\mathbf{\Lambda}) \mathbf{U}^{*}\mathbf{f} 
\stackrel{(\ref{eq:L})}{=}  \mathcal{P}(\mathbf{\mathcal{L}}) \mathbf{f}, 
\label{eq:filtering-b}
\end{align} 
thus, only requiring polynomial matrix operations on $\mathbf{\mathcal{L}}$. 

\subsection{Dataset}
The Human Connectome Project (HCP) dataset \cite{HCP} was used in this study. In particular, we use the 100 unrelated adult subject sub-group (54\% female, mean age = 29.11$\pm$ 3.67, age range = 22-36), which we denote as the HCP100 subject set. The study was approved by the Washington University Institutional Review Board and informed consent was obtained from all subjects.  
We use the minimally preprocessed structural data (0.7 mm$^{3}$) and task fMRI data (2 mm$^{3}$). 
Our proposed method heavily relies on the accurate co-registration between the structural and functional data from HCP. A full description of the imaging parameters and image prepocessing can be found in \cite{Glasser2013}. Task-based fMRI data for each subject consists of 1940 time frames, across seven functional tasks: Emotion, Gambling, Language, Motor, Relational, Social and Working Memory, including 23 experimental conditions in total; see labels in the bottom of Fig.~\ref{fig:results2}.



\subsection{Design of cerebral hemisphere cortex (CHC) graphs}

We design voxel-resolution graphs that encode the hemispheric cortical morphology, unique to each individual. 
The design is based on cortical ribbons, extracted using the FreeSurfer software package \cite{Fischl2012}. 
Ribbon files have been extensively leveraged to reliably measure the thickness of the gray matter of the human cerebral cortex and detect pathology induced variations of less than 0.25 mm \cite{Fischl2000}. 
The 0.7 mm$^3$ resolution ribbon file from an HCP subject is first downsampled to 1.25 mm$^3$. 
The resulting mask is morphologically processed to ensure that it is a single connected structure; i.e., any voxel within the ribbon that is not adjacent to another voxel within its 6-neighbourhood is removed. 
The remaining voxels are treated as preliminary vertices of the graph. 
The associated preliminary set of edges are defined based on the connectivity of voxels in 3D space similar to that in \cite{Behjat2015}: a vertex $i$ is considered connected to vertex $j$ if their associated voxels lie within each other's 26-neighborhood. 
The edges are treated as binary connections, and thus, they are assigned no weight. 

\begin{figure}[!]
\centering
\includegraphics[width=0.57\textwidth,trim={25cm 18.5cm 22cm 16.4cm},clip]{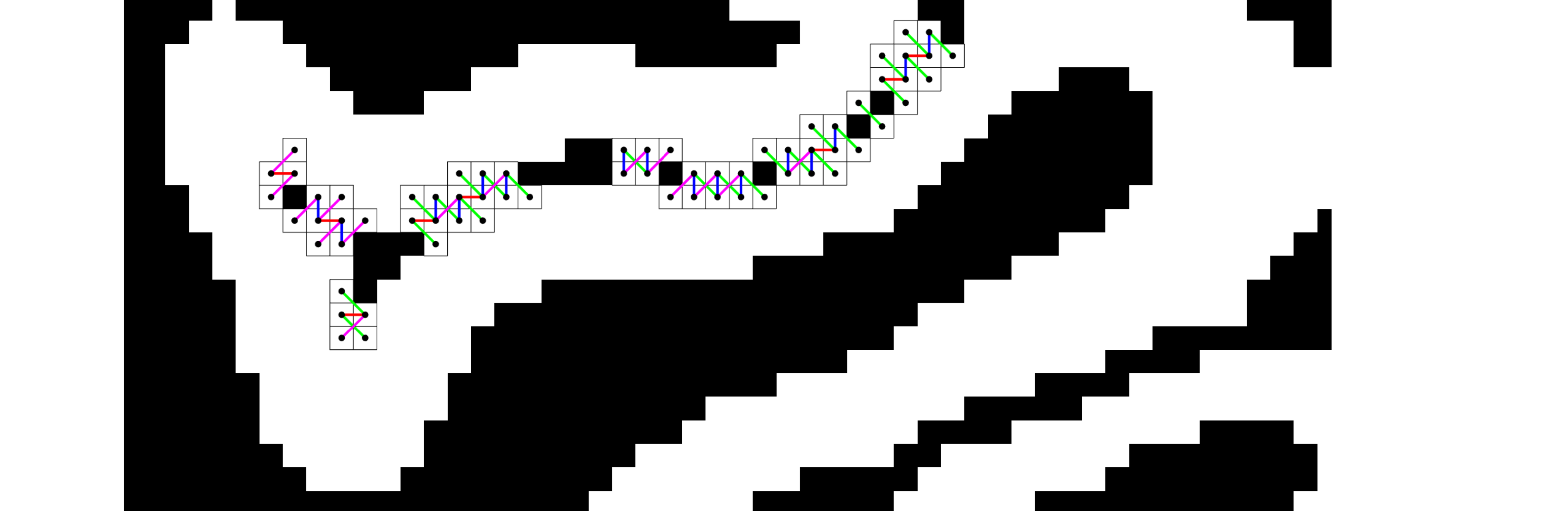}
\caption{
Graph edge pruning. 
Pruned graph edges are displayed, color-coded to differentiate horizontal, vertical and diagonal connections in the 2D plane. 
Voxels corresponding to the affected graph vertices are marked with a square.
} 
\label{fig:pruning}
\end{figure}

Due to limited voxel resolution, the extracted edges can consist of spurious connections that are not anatomically justifiable, for instance, at touching banks of sulci. 
By incorporating pial surface representations, any edge that has its ends on opposite sides of the pial surface is removed; we denote this procedure as ``pruning''. 
Figure~\ref{fig:pruning} illustrates an example set of pruned edges. 
Such pruned edges represent connections that encode only Euclidean adjacency rather than geodesic adjacency. 
Failure to remove these spurious edges would result in an inaccurate encoding of the cortical topology. 
Vertices from the preliminary set that become unconnected as a result of pruning are removed; the remaining vertices are treated as the graph's vertex set. 

\subsection{Spectral energy of fMRI data on CHC graphs}



Functional volumes from each subject were coregistered with the subject's ribbon file and resampled to 1.25 mm$^3$. Graph signals were then constructed by extracting values at voxels corresponding to the graph nodes. Due to sheer number of functional volumes, we find it beneficial to study the ensemble spectral properties of a set of signals, associated to either the same subject, task, or experimental condition. Let $\mathcal{F} = \{\tilde{\mathbf{f}}_{s}\}_{s=1}^{S}$ denote a set of $S$ graph signals defined on $\mathcal{G}$, where $\tilde{\mathbf{f}}_{s}$ denotes the de-meaned and normalized version of $\mathbf{f}_{s} \in \mathbb{R}^{N}$, obtained as $\tilde{\mathbf{f}}_{s} = (\mathbf{f}_{s} - \mathbf{u}_{1}^{\ast} \mathbf{f}_{s}\mathbf{u}_{1})/||\mathbf{f}_{s} - \mathbf{u}_{1}^{\ast} \mathbf{f}_{s}\mathbf{u}_{1}||_{2}^{2}$. Moreover, due to the voxel-resolution nature of CHC graphs, they are very large, with mean: $123 \times 10^{3} \pm 12 \times 10^{3}$ on the HCP100 subject set. This renders it challenging to study variations in spectral content of signals on CHC graphs at the resolution of individual eigenvalues. As such, we define a 
measure of spectral energy on $\mathcal{F}$ as 
\begin{equation}
\forall \lambda \in [0,\lambda_{\text{max}}],\quad E \left( \mathcal{F},\lambda \right) = \frac{1}{S}\sum_{s=1}^{S} \sum_{i=1}^{C(\lambda)} \left \vert \fHatTilde_{s}[i] \right \vert^2, 
\label{eq:E_ensemble}
\end{equation}  
where $C(\lambda)$ denotes the index of the largest eigenvalue smaller than lambda, i.e.,  
$
C(\lambda) = \argmin_{i \in 1,\cdots,N} \{\lambda-\lambda_i \}~\text{ s.t.}~\lambda-\lambda_i \ge0.
$

\begin{figure}[!]
\centering
\includegraphics[width=0.48\textwidth]{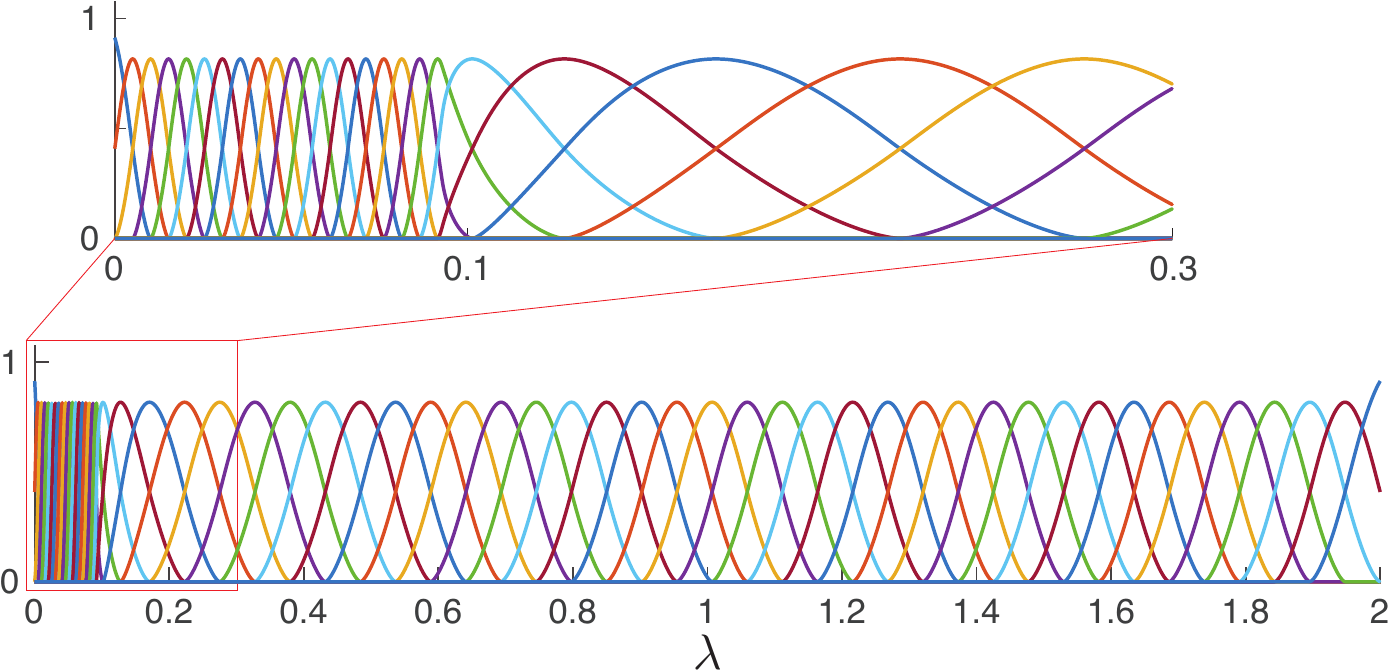} 
\caption{Spline-type system of spectral used for estimating spectral energy of cortical fMRI signals on CHC graphs. 
}
\label{fig:hbSosks}
\end{figure}

Computing $E \left( \mathcal{F},\lambda \right)$ for $\lambda$ at the upper parts of the spectrum is highly impractical due to their sheer size of CHC graphs. 
An alternative approach is to estimate the spectral content of $\mathcal{F}$ 
across a large set of subbands specified by a uniform system of spectral kernels associated to a tight frame \cite{Behjat2019chapter}. 
In this work, we exploit a spline-type system of spectral kernels, denoted $\{k_j(\lambda)\}_{j=1}^{J=57}$, see Fig.~\ref{fig:hbSosks}. 
The design provides a smooth transition from narrow band kernels at the lower-end of the spectrum, up to $\lambda=0.1$, to kernels with 10 fold wider spectral bands. The choice of $0.1$ is based on previous findings of the significance of the very low spectral end 
in capturing substantial energy content of fMRI data on similarly designed gray matter brain graphs \cite{Behjat2015, Behjat2016}. It can be shown that $\{k_j(\lambda)\}_{j=1}^{J}$ satisfy the tight Parseval frame property, 
i.e., $ \forall \lambda \in [0,\lambda_{\text{max}}], \quad  T(\lambda) = \sum_{j=1}^{J} \vert k_{j} (\lambda)\vert^2 = 1$, which guarantees energy conservation between the vertex and spectral domain, i.e., $\forall \mathbf{f} \in \mathbb{R}^{N}, \quad \sum_{j=1}^{J} || k_{j}(\mathbf{\mathcal{L}}) \mathbf{f}||_2^{2} = ||\mathbf{f}||_2^{2}$. We defer detailed description of the design to future work. 
We define a coarse measure of spectral energy on $\mathcal{F}$ as 
\begin{equation}
E\left(\mathcal{F},c_j\right) =    \frac{1}{S}\sum_{s=1}^{S} \sum_{i=1}^{j}|| k_{i}(\mathbf{\mathcal{L}}) \mathbf{f}||_2^{2}, \quad j = 1,\ldots, J,
\label{eq:Ek}
\end{equation}
where $c_j$ denotes the center of mass of $k_j(\lambda)$ for $ j=1,\ldots,J-1$, and $c_{J} =2$. 
We implement (\ref{eq:Ek}) using the polynomial approximation scheme, cf. (\ref{eq:filtering-b}). In particular, we approximate $\{k_j(\lambda)\}_{j=1}^{J}$ with Chebyshev polynomials of varying degree, mean: 300 $\pm$ 200, denoted $\{\tilde{k}_j(\lambda)\}_{j=1}^{J}$, such that $\forall \lambda \in [0, \lambda_{\text{max}}], \vert  \sum_{j=1}^{J} \vert \tilde{k}_{j} (\lambda)\vert^2 - T(\lambda)| \le 0.01$. 

\section{Results}
For each subject, regressors associated to each of the 23 experimental condition were obtained by convolving the associated paradigms with the hemodynamic response function using SPM12 \cite{spm12}. 
To study fMRI signal components related to the experimental tasks, subsets of the fMRI time frames were extracted; specifically, fMRI volumes at time instances associated to regressor values of 0.8 and greater were extracted. 
CHC graph signals were extracted from each volume and used to define 30 subject-specific signal sets $\mathcal{F}$: one set associated to each experimental condition, and one set associated to each task obtained as the union of the task's experimental condition sets. Functional data on the left hemispheres were studied.

\begin{figure}[!]
\centering
\includegraphics[width=0.48\textwidth]{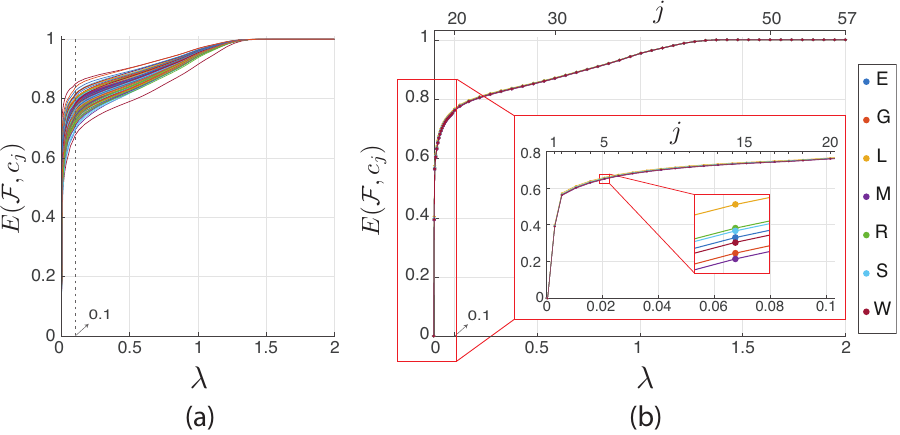} 
\caption{
(a) Cumulative spectral energy of the graph signal sets associated to the Emotion task, across 100 subjects. 
(b) Cumulative spectral energy of the graph signal sets associated to the seven tasks, averaged across 100 subjects; E: emotion, L: language, R: relational, G: gambling, M: motor, S: social, W: working memory.
}
\label{fig:results1a}
\end{figure}

Fig.~\ref{fig:results1a}(a) shows estimates of the energy spectral density for the Emotion task across subjects, 
reflecting that a substantial amount of energy content is contained within a very narrow lower-end spectral band, below $\lambda=0.1$. 
Fig.~\ref{fig:results1a}(b) shows ensemble average of the curves shown in  Fig.~\ref{fig:results1a}(a), as well for the remaining six tasks, manifesting that the ensemble energy densities minimally deviate between tasks. 

\begin{figure}[!]
\centering
\includegraphics[width=0.48\textwidth]{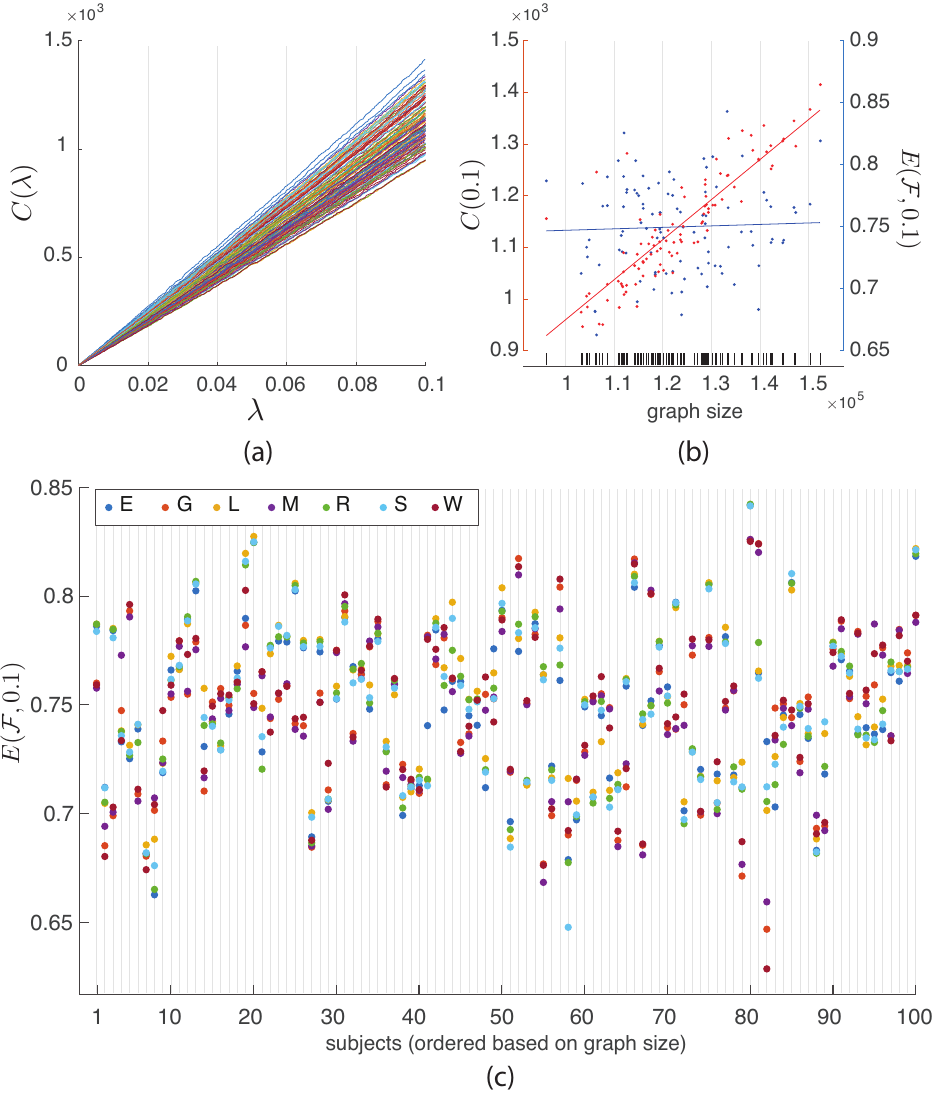} 
\caption{
(a) Distribution of eigenvalues in the lower-end $[0,0.1]$ of the graph Laplacian spectra across 100 subjects. 
(b) Relation between CHC graph size, $C(0.1)$ and $E(\mathcal{F},0.1)$ for the Emotion task. 
The black vertical bars on the horizontal axis mark the size of each subjects graph. 
(c) Comparison of $E(\mathcal{F},0.1)$ across subjects and tasks. The scatter plot for $E(\mathcal{F},0.1)$ of the Emotion task is replicated from (b), but horizontally rearranged to prevent overlapping dots between subjects. In several subjects, $E(\mathcal{F},0.1)$ becomes very similar for multiple tasks, and thus, the tasks cannot be visually distinguished due to the overlap.
}
\label{fig:results1}
\end{figure}

Fig.~\ref{fig:results1}(a) shows the cumulative distribution of eigenvalues in the lower-end $[0,0.1]$ spectra of CHC graphs, across subjects. 
The observed variation has been shown to encode morphological differences between cortical hemispheres across subjects \cite{Maghsadhagh2019}. 
Fig.~\ref{fig:results1}(b) shows that the size of CHC graphs and their associated $C(0.1)$ are highly correlated (correlation coefficient 0.86, 95\% confidence interval 0.80-0.91), whereas negligable correlation is observed between $C(0.1)$ and $E(\mathcal{F},0.1)$ (correlation coefficient 0.05, 95\% confidence interval -0.15-0.24). 
Variations in $E(\mathcal{F},0.1)$ are thus not a mere reflection of variations in the spectra of CHC graphs. 
Fig.~\ref{fig:results1}(c) shows the distribution of $E(\mathcal{F},0.1)$ across subjects and tasks. 
$E(\mathcal{F},0.1)$ captures a large extend of variation across subjects, as well as under different functional load within the same subject, see vertical distribution of the seven colored dots.
As the subjects are ordered based on their CHC graph size, the negligible correlation between $E(\mathcal{F},\lambda)$ and graph size can be visually inferred for all seven tasks; 
fitted lines are not plotted as they were almost horizontal, similar to that observed for the Emotion task in Fig.~\ref{fig:results1}(b).

\begin{figure}[t]
\centering
\includegraphics[width=0.47\textwidth]{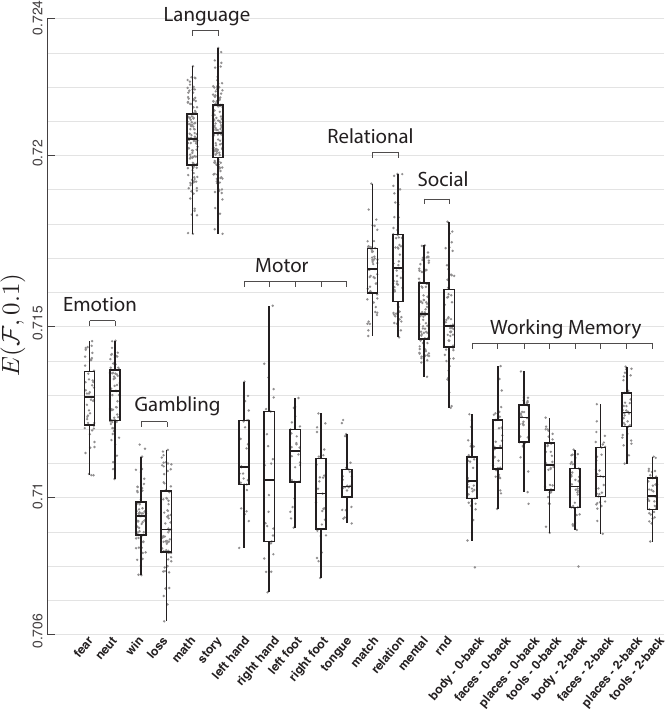} 
\caption{
Comparison of $E(\mathcal{F},\lambda)$ across 23 experimental conditions within seven tasks, for a single subject, corresponding to subject 40 in Fig.~\ref{fig:results1}(c). Each dot is associated to a single CHC graph signal, and thus, a single fMRI volume, i.e., a single time instance. The number of samples varies within each condition due to variations in the time length of associated experimental paradigms. 
} 
\label{fig:results2}
\end{figure}

Results in Fig.~\ref{fig:results1}(c) reflect noticeable variations at the subject level, across the seven tasks. 
To further explore this within-subject and task specific variation, the proposed spectral energy measure was studied across the 23 experimental conditions. 
Fig.~\ref{fig:results2} shows the results on a single subject, 
corresponding to subject 40 in Fig.~\ref{fig:results1}(c). 
Not only is variation observed between tasks, but more interestingly, between different experimental conditions within tasks, in particular, for the Motor and Working Memory tasks. 
These results further signify the capability of the eigenvectors of CHC graphs in capturing anatomically-constrained functional variations. 



\section{Conclusions}
We proposed an approach to study fMRI data on volumetric, voxel-resolution CHC graphs that encode local cortical geometry of a hemisphere. 
Compared to cortical surface mapping schemes, the proposed approach maintains the analysis within the native volumetric space, and thus alleviates the need to project 3-D volumes on to 2-D cortical surfaces. 
Experimental results signified the capability of CHC graphs' eigenmodes in capturing task specific spatial patterns of brain activity. 
Our future work will be directed towards: 
i) extending the results to larger cohorts of subjects and to white matter \cite{Tarun2019isbi,Abramian2019}, 
ii) investigating potential extension of the method by using localized spectral energy metrics \cite{Stankovic2017} and fast GFT \cite{Magoarou2017}, 
and iii) implementing fMRI filtering using scalable critically sampled graph filter banks \cite{Shuman2019scalable} to overcome limitations due to the sheer size of voxel-resolution CHC graphs. 

\section{Acknowledgements}
Data were provided by the Human Connectome Project, WU-Minn Consortium (Principal Investigators: David Van Essen and Kamil Ugurbil; 1U54MH091657) funded by the 16 NIH Institutes and Centers that support the NIH Blueprint for Neuroscience Research; and by the McDonnell Center for Systems Neuroscience at Washington University. The computations were performed in part on resources provided by the Swedish National Infrastructure for Computing (SNIC) through the LUNARC center at Lund University, under Project SNIC 2019/6-36. 

\clearpage
\bibliographystyle{IEEEbib}
\bibliography{hb}

\end{document}